\documentclass[review]{elsarticle}

\usepackage{lineno, hyperref}
\usepackage{amsmath, amssymb}
\usepackage{graphicx}
\usepackage{mathtools}

\journal{Journal of \LaTeX\ Templates}

\begin{document}

\begin{frontmatter}

\title{Exact bounds for dynamical critical exponents of transverse-field Ising chains with a correlated disorder}
\author{Tatsuhiko Shirai\corref{correspondingauthor}}
\address{Department of Computer Science and Communications Engineering, Waseda University
3-4-1, Ookubo, Shinjuku-ku, Tokyo, 169-8555 Japan}
\cortext[correspondingauthor]{Corresponding author}
\ead{tatsuhiko.shirai@aoni.waseda.jp}

\author{Shu Tanaka\corref{secondauthor}}
\address{Department of Applied Physics and Physico-Informatics, Keio University, 3-14-1, Hiyoshi, Kohoku-ku, Yokohama, 223-8522, Japan}
\address{Green Computing System Research Organization, Waseda University, Wasedamachi-27, Shinjuku-ku, Tokyo, 162-0042, Japan}
\ead{shu.tanaka@appi.keio.ac.jp}

\begin{abstract}
This study investigates the dynamical critical exponent of disordered Ising chains under transverse fields to examine the effect of a correlated disorder on quantum phase transitions.
The correlated disorder, where the on-site transverse field depends on the nearest-neighbor coupling strengths connecting the site, gives a qualitatively different result from the uncorrelated disorder.
In the uncorrelated disorder cases where the transverse field is either homogeneous over sites or random independently of the nearest-neighbor coupling strengths, the dynamical critical exponent is infinite.
In contrast, in the presence of the correlated disorder, we analytically show that the dynamical critical exponent is finite.
We also show that the dynamical critical exponent depends on the tuning process of the transverse field strengths.
\end{abstract}


\end{frontmatter}


\section{Introduction}\label{intro}
Quantum phase transitions (QPTs) have attracted much interest in condensed matter physics and statistical physics~\cite{sachdev2007quantum}.
A QPT occurs at absolute zero temperature as a parameter other than the temperature is varied.
It is accompanied with a qualitative change in the properties of the ground state.
Critical exponents describe the second-order (continuous) QPTs with an argument based on universality.
The dynamical critical exponent $z$ determines the relation between the characteristic energy scale and the length scale around the critical point, i.e., $\Delta \sim \xi^{-z}$, where $\Delta$ and $\xi$ are the energy gap between the ground state and the first-excited state and the correlation length in the ground state, respectively.

A disorder-free Ising chain with homogeneous transverse fields is a simple model for studying the properties of the second-order QPTs~\cite{sachdev2007quantum, tanaka2017quantum, suzuki2012quantum}.
The Hamiltonian of the $N$-spin chain is given by
\begin{equation}
    H=-\sum_{i=1}^{N-1} J_i \sigma_i^z \sigma_{i+1}^z -\sum_{i=1}^N \Gamma_i \sigma_i^x,
    \label{eq:model_intro}
\end{equation}
where $\vec{\sigma}_i=(\sigma_i^x, \sigma_i^y, \sigma_i^z)$ are Pauli spin matrices, and nearest-neighbor couplings $\{J_i\}$ and transverse fields $\{\Gamma_i\}$ are site independent (i.e., $J_i=J$ and $\Gamma_i=\Gamma$).
As $\Gamma$ decreases, the system shows a continuous QPT from a paramagnetic phase to an ordered phase.
The various properties of the QPT are exactly obtained~\cite{pierre1970one}, e.g. the dynamical critical exponent $z$ is given by $z=1$~\cite{sachdev2007quantum}.
This model still gives a new concept on physics by showing a close relation to the model with a topological order~\cite{feng2007topological}.

Then, consider a disordered Ising chain with transverse fields.
This model is of interest for the study of the relation between the properties of the QPTs and disorder.
In the Hamiltonian in eq.~(\ref{eq:model_intro}),
the nearest-neighbor coupling strengths $J_i$ are randomly distributed.
The transverse field is either set as homogeneous over sites, $\Gamma_i=\Gamma$~\cite{dziarmaga2006dynamics}, or randomly chosen independent of the nearest-neighbor coupling strengths~\cite{ceneva2007adiabatic,suzuki2011kibble}.
In both cases, the transverse fields are uncorrelated with the nearest-neighbor couplings and qualitatively similar results have been reported.
As well as the case without disorder, the system shows a continuous QPT between a paramagnetic phase and an ordered phase.
However, the properties of the QPT are different from the disorder-free case.
Around the critical point, the Griffiths--McCoy (GM) singularity arises~\cite{griffiths1969nonanalytic, mccoy1969incompleteness} due to the presence of a statistically rare region where the spins are strongly coupled in the paramagnetic phase or decoupled in the ordered phase.
The GM singularity leads to a continuously varying dynamical exponent in the paramagnetic phase, resulting in the divergence of susceptibility for a finite range of the transverse field strength~\cite{igloi1998anomalous, igloi2001griffiths, igloi2002exact}.
At the critical point, the dynamical critical exponent is infinite, $z=\infty$.
This model has been investigated by mapping to a system with free fermions~\cite{young1996numerical} and real-space renormalization group analysis~\cite{fisher1995critical, fisher1992random}.

Here, we investigate the effect of a correlated disorder on the QPT in the transverse-field Ising chain.
We set the transverse field strength on each site according to the coupling strengths connecting the site.
Namely, $\Gamma_i$ depends on $J_{i-1}$ and $J_i$ in eq.~(\ref{eq:model_intro}).
Unlike the uncorrelated disorder cases, we show that $z$ is finite in this model.
This study shows that $z$ depends on the distribution of the nearest-neighbor coupling strength: the weak-disorder case and the strong-disorder case.
We analytically show that $z=1$ in the weak-disorder case and $\max(D(1/2+|s-1/2|)+1/2, 1)\leq z \leq D+1$ in the strong-disorder case, where $s \in [0,1]$ is a parameter for tuning the transverse fields and $D$ is a parameter for the disorder strength.
We numerically estimate $z$ in the strong-disorder case, and reveal that $z$ depends on the process to tune the transverse field strength.

The transverse-field Ising chain with the correlated disorder was investigated in previous studies.
In weak-disorder case, $z=1$ was obtained in analysis based on an exact diagonalization method~\cite{hoyos2011protecting, knysh2020quantum}.
In strong-disorder case, $D$-dependence of $z$ was obtained by using an exact diagonalization method~\cite{hoyos2011protecting} and a strong-disorder renormalization group analysis~\cite{getelina2016entanglement}.
Our study should be distinguished from these numerical studies since exact results on $z$ are analytically obtained.

The rest of this paper is organized as follows.
Section~\ref{sec:model} introduces the model of the transverse-field Ising chains with the correlated disorder.
Section~\ref{sec:bound} gives both analytic and numerical results of the dynamical critical exponents.
Section~\ref{sec:conclusion} concludes our study and provides a future direction.
\ref{appendixA} shows the mapping details from a spin system to a free fermion system.
\ref{appendixB} gives the numerical study of the correlation length around the critical point.

\section{Model}\label{sec:model}
This study investigates the effect of a correlated disorder on the properties of QPTs through a prototypical model. 
Thus, we consider an Ising model on a one-dimensional chain of $N$ spins under transverse fields.
The Hamiltonian is given in the form of eq.~(\ref{eq:model_intro}).
%
%
The nearest-neighbour coupling strengths are assumed to be ferromagnetic without loss of generality since the sign of $J_i$ can be changed by applying a gauge transformation.
The gauge transformation consists of a $\pi$ rotation around the $x$-axis in the spin space as $\sigma_i^x \to \sigma_i^x$, $\sigma_i^y \to -\sigma_i^y$ and $\sigma_i^z \to -\sigma_i^z$, which flips the sign of $\sigma_{i-1}^z \sigma_i^z$ term and $\sigma_i^z \sigma_{i+1}^z$ term.
The coupling strengths are chosen from two different types of distribution: weak disorder and strong disorder.
The distribution in the weak disorder case is uniform over $(J^{(0)}, 1]$ and gapped (i.e., $J^{(0)} > 0$).
That is, 
\begin{equation}
    \pi_{\rm w}(J_i)=\left\{
    \begin{aligned}
    &(1-J^{(0)})^{-1} \text{ for } J^{(0)} < J_i \leq 1,\\
   &\quad 0 \quad\qquad \text{ otherwise.}
    \end{aligned}
    \right.
    \label{distribution_weak}
\end{equation}
The distribution in the strong-disorder case is a gapless power-law distribution over $(0,1]$ given as
\begin{equation}
    \pi_{\rm s}(J_i)=\left\{
    \begin{aligned}
    &\frac{1}{D} J_i^{-1+\frac{1}{D}} \text{ for } 0 < J_i \leq 1,\\
   &\quad 0 \quad\qquad \text{ otherwise.}
    \end{aligned}
    \right.
    \label{distribution_weak}
\end{equation}
$D$ is a positive real value and denotes the disorder strength~\cite{hoyos2011protecting}.

%
In the correlated disorder case, the transverse field strengths are given as
\begin{equation}
    \left\{
    \begin{aligned}
    \ln \frac{\Gamma_1}{\Gamma} &= (1-s) \ln J_{1},\\
    \ln \frac{\Gamma_i}{\Gamma} &= s \ln J_{i-1} + (1-s) \ln J_{i} \quad \text{ for } 2\leq i \leq N-1,\\
    \ln \frac{\Gamma_N}{\Gamma} &= s \ln J_{N-1},
    \end{aligned}
    \right.
    \label{tuned}
\end{equation}
where $s$ is a real value in the interval $[0,1]$.
$s$ is a parameter for tuning the transverse fields.
The transverse field strength at site $i$ depends on the coupling strengths, $J_{i-1}$ and $J_i$.
The overall strengths are controlled by $\Gamma$.

Let $\Gamma_{\rm c}$ be the critical point.
The ground state of $H$ in eq.~(\ref{eq:model_intro}) is the paramagnetic phase when $\Gamma>\Gamma_{\mathrm{c}}$ and the ordered phase when $\Gamma < \Gamma_{\mathrm{c}}$.
$\Gamma_{\rm c}$ is determined by~\cite{fisher1995critical,shankar1987nearest}
\begin{equation}
    \Delta_J = \Delta_\Gamma,
    \label{critical}
\end{equation}
where $\Delta_J$ and $\Delta_\Gamma$ are the averages of $\ln J_i$ and $\ln \Gamma_i$ in the bulk (i.e., $2\leq i \leq N-1$) over disorder.
$\Delta_J$ is defined as
\begin{equation}
    \Delta_J \coloneqq  \int_{-\infty}^\infty dJ_1 \cdots \int_{-\infty}^\infty dJ_{N-1} \left( \prod_{j=1}^{N-1} \pi_{\rm s/w}(J_j) \right) \ln J_i.
    \label{DeltaJ}
\end{equation}
In the correlated disorder case, since $\Gamma_i$ is determined by $J_{i-1}$ and $J_i$, $\Delta_\Gamma$ is defined as
\begin{equation}
    \Delta_\Gamma \coloneqq  \int_{-\infty}^\infty dJ_1 \cdots \int_{-\infty}^\infty dJ_{N-1} \left( \prod_{j=1}^{N-1} \pi_{\rm s/w}(J_j) \right) \ln \Gamma_i.
    \label{DeltaG}
\end{equation}
Substituting eqs.~(\ref{DeltaJ}) and~(\ref{DeltaG}) into eq.~(\ref{critical}) gives $\Gamma_{\mathrm{c}}=1$ for any value of $s$ in both weak-disorder case and strong-disorder case.
Below, the average of a random variable $O$ over disorder is denoted by
\begin{equation}
    [O]_{\rm ave}=\int_{-\infty}^\infty dJ_1 \cdots \int_{-\infty}^\infty dJ_{N-1} \left( \prod_{j=1}^{N-1} \pi_{\rm s/w}(J_j) \right) O.
\end{equation}

\section{Dynamical critical exponent}\label{sec:bound}
Here, we calculate dynamical critical exponents $z$ in the weak-disorder case and the strong-disorder case.
The energy scale of the system is characterized by the energy gap between the ground state and the first-excited state, $\Delta (\{J_i \})$.
The probability distribution function (PDF) of the energy gap $\Delta$ at $\Gamma$ and $N$ is given by
\begin{equation}
\mathcal{P}(\Delta; \Gamma, N)=[\delta(\Delta-\Delta (\{J_i \}))]_{\rm ave},
\end{equation}
where $\delta(\cdot)$ is a delta function.

Around the critical point, the PDF in the limit of $N \to \infty$ obeys the scaling form~\cite{sachdev2007quantum},
\begin{equation}
     \mathcal{P}(\Delta; \Gamma, N\to \infty) \sim \xi^z \mathcal{P}_{(1)}(\Delta \xi^z)
\end{equation}
where $\mathcal{P}_{(1)}(\cdot)$ is a PDF.
Here, $\xi$ is the average correlation length~\cite{hoyos2011protecting} (see~\ref{appendixB} for the definition of $\xi$ and the comparison to another correlation length referred to as the typical correlation length).
The diverging length scale is described as
\begin{equation}
\xi \sim |\Gamma -\Gamma_{\mathrm{c}}|^{-\nu},
\end{equation}
where $\nu$ is the critical exponent.
Thus, the PDF is rewritten as
\begin{equation}
     \mathcal{P}(\Delta; \Gamma, N\to \infty) \sim |\Gamma -\Gamma_{\mathrm{c}}|^{-z \nu} \mathcal{P}_{(1)}(\Delta |\Gamma -\Gamma_{\mathrm{c}}|^{-z \nu}).
\end{equation}

The finite-size scaling~\cite{cardy1996scaling} gives
\begin{align}
     \mathcal{P}(\Delta; \Gamma, N) &\sim |\Gamma -\Gamma_{\mathrm{c}}|^{-z \nu} \mathcal{P}_{(2)}(\Delta |\Gamma -\Gamma_{\mathrm{c}}|^{-z \nu}, N \xi^{-1}) \nonumber\\
     &\sim |\Gamma -\Gamma_{\mathrm{c}}|^{-z \nu} \mathcal{P}_{(2)}(\Delta |\Gamma -\Gamma_{\mathrm{c}}|^{-z \nu}, N|\Gamma-\Gamma_{\rm c}|^{\nu})\nonumber\\
     &\sim N^z \mathcal{P}_{(3)}(\Delta N^z, N^{\frac{1}{\nu}}|\Gamma-\Gamma_{\rm c}|)
\end{align}
where $\mathcal{P}_{(2)}(\cdot, \cdot)$ and $\mathcal{P}_{(3)}(\cdot, \cdot)$ are PDFs.
As a result, at the critical point, we find
\begin{equation}
    \mathcal{P}(\Delta; \Gamma_{\rm c}, N) \sim N^z \mathcal{P}_{(3)}(\Delta N^z, 0).
    \label{probability}
\end{equation}
Thus, $z$ is obtained by the system-size dependence of the  probability distribution function of the energy gap at the critical point.
Instead of the PDF, we consider the moment of $\Delta$.
Equation~(\ref{probability}) implies that the $m$-th moment ($m \in \mathbb{N}$) obeys
\begin{align}
    [\Delta^m]_{\rm av} &= \int \Delta^m \mathcal{P}(\Delta; \Gamma_{\rm c}, N) d\Delta,\nonumber\\
    &\sim \int \Delta^m \mathcal{P}_{(3)}(\Delta N^z, 0) d(\Delta N^z)\sim N^{-mz}
    \label{eq:delta_vs_N}
\end{align}

In subsection~\ref{subsec:analytic}, we analytically obtain a lower bound and an upper bound of $[\Delta^m]_{\rm av}$.
We show from the bounds that $z=1$ in the weak-disorder case and $\max (D(1/2+|s-1/2|)+1/2, 1)\leq z \leq D+1$ in the strong-disorder case.
In subsection~\ref{subsec:numerics}, we numerically estimate $z$ in the strong-disorder case.

\subsection{Analytic results}\label{subsec:analytic}
This subsection provides the analytic results of $z$.
The derivation of the results is valid for the free boundary condition.
Since the relation between the $m$-th moment of the energy gap $[\Delta^m]_{\mathrm{av}}$ and the dynamical critical exponent $z$ is given by eq.~(\ref{eq:delta_vs_N}), an upper bound and a lower bound of $z$ are obtained by inequalities between $[\Delta^m]_{\mathrm{av}}$ and the system size $N$. More precisely, an upper/lower bound of $z$ is obtained from a lower/upper bound of $[\Delta^m]_{\mathrm{av}}$~\footnote{
Estimating $z$ only through $[\Delta]_{\rm ave}$ could be problematic, since $[\Delta]_{\rm ave}$ could be dominated by some large $O(1)$ values of rare samples.
The higher-order moments of $\Delta$ are required to determine $z$.}. The details are given below.

Mapping to a fermion system relates $\Delta$ in the following eigenvalue problem
\begin{equation}
    \mathbf{M}\vec{\phi}_n=\Lambda_n^2 \vec{\phi}_n.
\end{equation}
For the specific form of the matrix $\mathbf{M}$, see eq.~(\ref{eigenvalue}) in~\ref{appendixA}.
When the eigenvalues $\Lambda_n (\geq 0)$ are arranged in ascending order (i.e., $\Lambda_1 \leq \cdots \leq \Lambda_N$), the energy gap is given by
\begin{equation}
    \Delta = \Lambda_1.
\end{equation}

By tuning the transverse field strengths in eq.~(\ref{tuned}), matrix $\mathbf{M}$ is written as
\begin{equation}
    \mathbf{M}=4 (\mathbf{J})^s \mathbf{L} (\mathbf{J}_+)^{2(1-s)} \mathbf{R} (\mathbf{J})^s,
    \label{matrixM}
\end{equation}
where $\mathbf{J}$ and $\mathbf{J}_+$ are diagonal matrices
\begin{equation}
    [\mathbf{J}]_{ij}=\left\{
    \begin{aligned}
    1 \qquad &\text{for } i=j=1,\\
    J_{i-1} \qquad &\text{for } 2\leq i=j \leq N,\\
    0 \qquad &\text{otherwise},
    \end{aligned}
    \right.
\end{equation}
\begin{equation}
    [\mathbf{J}_+]_{ij}=\left\{
    \begin{aligned}
    J_i \qquad &\text{for } 1\leq i=j \leq N-1,\\
    1 \qquad &\text{for } i=j=N,\\
    0 \qquad &\text{otherwise},
    \end{aligned}
    \right.
\end{equation}
and
\begin{equation}
    [\mathbf{L}]_{ij}=\left\{
    \begin{aligned}
    \Gamma \qquad &\text{for } i=j,\\
    1 \qquad &\text{for } i-j=1,\\
    0 \qquad &\text{otherwise},
    \end{aligned}
    \right.
\end{equation}
and $\mathbf{R}=\mathbf{L}^\dagger$.
Here, $[\mathbf{J}]_{ij}$, $[\mathbf{J}_+]_{ij}$, and $[\mathbf{L}]_{ij}$ are the $(i,j)$-th element of matrices $\mathbf{J}$, $\mathbf{J}_+$, and $\mathbf{L}$, respectively.
Below, we denote the $(i,j)$-th element of matrix $\mathbf{C}$ by $[\mathbf{C}]_{ij}$.

First, we provide a lower bound of the energy gap.
Consider the following inequality
\begin{equation}
    \mathbf{M} \geq \mathbf{M}_0 \coloneqq 4(J^{(1)})^2 \mathbf{LR}, \label{inequality}
\end{equation}
where $J^{(1)}=\min \{J_i\}_{i=1}^{N-1}$.
Herein, for Hermite matrices $\mathbf{H}$ and $\mathbf{G}$, $\mathbf{G}\leq \mathbf{H}$ means that $\mathbf{H}-\mathbf{G}$ is positive semidefinite.
This inequality is derived as follows.
We find 
\begin{equation}
    \mathbf{M}= \mathbf{M}_0+\mathbf{M}_1+\mathbf{M}_2+\mathbf{M}_3,
\end{equation}
where
\begin{equation}
    \left\{    
    \begin{aligned}
    \mathbf{M}_1&=4\mathbf{J}^s \mathbf{L} [\mathbf{J}_+^{2(1-s)}-(J^{(1)})^{2(1-s)} \mathbf{I}] \mathbf{R} \mathbf{J}^s,\\
    \mathbf{M}_2&=4(J^{(1)})^{2(1-s)}[\mathbf{J}^s - (J^{(1)})^{s} \mathbf{I}] \mathbf{L} \mathbf{R} \mathbf{J}^s,\\
    \mathbf{M}_3&=4(J^{(1)})^{2-s} \mathbf{LR} [\mathbf{J}^s - (J^{(1)})^{s} \mathbf{I}].
    \end{aligned}
    \right.
\end{equation}
Here, $\mathbf{I}$ denotes the $N \times N$ identity matrix.
$\mathbf{M}_1$ is positive semi-definite.
$\mathbf{M}_2$ and $\mathbf{M}_3$ are positive semi-definite, since the energy spectra of $\mathbf{M}_2$ and $\mathbf{M}_3$ are respectively identical to those of $\mathbf{M}_2'$ and $\mathbf{M}_3'$ given by
\begin{equation}
    \left\{  
    \begin{aligned}
    \mathbf{M}_2'=&4(J^{(1)})^{2(1-s)} \mathbf{R} \mathbf{J}^s [\mathbf{J}^s - (J^{(1)})^{s} \mathbf{I}] \mathbf{L},\\
    \mathbf{M}_3'=&4(J^{(1)})^{2-s} \mathbf{R} [\mathbf{J}^s - (J^{(1)})^{s} \mathbf{I}]\mathbf{L}.
    \end{aligned}
    \right.
\end{equation}
Because the sum of positive semi-definite matrices is positive semi-definite, we obtain $\mathbf{M}_1+\mathbf{M}_2+\mathbf{M}_3\geq 0$ and eq.~(\ref{inequality}).

Then, we introduce matrix $\mathbf{T}(t, u, \Gamma)$ whose matrix elements are given by
\begin{equation}
    [\mathbf{T}(t, u, \Gamma)]_{ij}=\left\{    
    \begin{aligned}
    &t+\Gamma^2 \text{ for } i=j=1,\\
    &1+\Gamma^2 \text{ for } 2\leq i=j \leq N-1,\\
    &u+\Gamma^2 \text{ for } i=j=N,\\
    &\Gamma \text{ for } |i-j|=1,\\
    &0 \text{ otherwise},
    \end{aligned}
    \right.
\end{equation}
where $t,u \in \mathbb{R}$.
The eigenvalues of $\mathbf{T}(t, u, \Gamma)$ are denoted by $\{\epsilon_n (t, u, \Gamma)\}_{n=1}^N$.
The matrix is related to matrices $\mathbf{LR}$ and $\mathbf{RL}$ as
\begin{equation}
    \mathbf{T}(0,1, \Gamma)=\mathbf{LR}, \text{ and } \mathbf{T}(1,0, \Gamma)=\mathbf{RL},
\end{equation}
respectively.
The matrix satisfies the inequality
\begin{equation}
    \mathbf{T}(t_1,u_1, \Gamma) \leq \mathbf{T}(t_2, u_2, \Gamma) \text{ for } t_1 \leq t_2 \text{ and } u_1 \leq u_2.
    \label{inequalityT}
\end{equation}

Weyl's monotonicity theorem~\cite{bhatia2013matrix} gives the lower bound from the inequality in eq.~(\ref{inequality}).
The eigenvalues of $\mathbf{M}_0$ are given by $4 (J^{(1)})^2 \epsilon_n(0,1,\Gamma)$.
The theorem gives
\begin{equation}
    \Lambda_1^2 \geq 4 (J^{(1)})^2 \min_n \epsilon_n(0,1, \Gamma),
\end{equation}
from which the lower bound of the energy gap is obtained as
\begin{equation}
    \Delta = \Lambda_1 \geq 2 J^{(1)} \sqrt{\min_n \epsilon_n(0,1, \Gamma)}.
    \label{lower_bound}
\end{equation}

We estimate an upper bound of $z$ using eq.~(\ref{lower_bound}).
At the critical point, $\Gamma=\Gamma_{\mathrm{c}}=1$, the eigenvalues of $\mathbf{T}(0,1, \Gamma_{\mathrm{c}})$ are given by
\begin{equation}
    \epsilon_n(0,1, \Gamma_{\mathrm{c}})=2+2 \cos  \frac{(n-\frac{1}{2})\pi}{N+\frac{1}{2}}.
\end{equation}
The eigenvalue is the smallest when $n=N$, and hence eq.~(\ref{lower_bound}) gives
\begin{equation}
    \Delta = \Lambda_1 \geq 2 J^{(1)} \sqrt{\epsilon_N(0,1, \Gamma_{\mathrm{c}})}=2 J^{(1)} \sqrt{2-2 \cos \frac{\pi}{N+\frac{1}{2}} }.
    \label{lower_bound2}
\end{equation}

In the weak-disorder case, since $J^{(1)}\geq J^{(0)}(>0)$, the lower bound of $\Delta$ [eq.~(\ref{lower_bound2})] gives for $m \in \mathbf{N}$
\begin{equation}
    [\Delta^m]_{\mathrm{av}} \geq (2 J^{(0)})^m \left(2-2 \cos \frac{\pi}{N+\frac{1}{2}}\right)^{\frac{m}{2}} \sim N^{-m}.
    \label{lower_bound_weak}
\end{equation}
This indicates $z\leq 1$ as an upper bound of $z$ in the weak-disorder case.

In the strong-disorder case, the lower bound of $\Delta$ [eq.~(\ref{lower_bound2})] gives for $m \in \mathbf{N}$
\begin{equation}
    [\Delta^m]_{\mathrm{av}} \geq 2^m [(J^{(1)})^m]_{\mathrm{av}} \left(2-2 \cos \frac{\pi}{N+\frac{1}{2}}\right)^{\frac{m}{2}}.
\end{equation}
The average of $(J^{(1)})^m$ over disorder is given by
\begin{align}
    [(J^{(1)})^m]_{\mathrm{av}}=&(N-1)! \int_0^1 dJ_1 \int_{J_1}^1 dJ_2 \cdots \int_{J_{N-2}}^1 dJ_{N-1} \left[ \prod_{i=1}^{N-1} \pi_{\rm s} (J_i)\right] J_1^m, \nonumber\\
    =& (N-1) B(mD+1, N-1),
\end{align}
where $B(\cdot, \cdot)$ is the beta function.
The beta function for $N \ge mD+1$ is bounded by
\begin{align}
    \ln [(N-1) B(mD+1, N-1)]=&-\sum_{n=1}^{N-1} \ln \left( 1+\frac{mD}{n}\right),\nonumber\\
    \geq &-\ln (mD+1)  -\int_1^{N-1} \ln \left(1+\frac{mD}{x}\right) dx,\nonumber\\
    = & mD \ln \frac{mD+1}{N-1} -(N+mD-1)\ln \left(1+\frac{mD}{N-1}\right),\nonumber\\
    \geq & mD \ln \frac{mD+1}{N-1} -2mD.
\end{align}
From the third line to the fourth line, we have used $\ln (1+x) \leq x$ and $N \geq mD+1$.
Hence, for $N \geq mD+1$
\begin{align}
    [\Delta^m]_{\mathrm{av}} \geq& 2^m (N-1) B(mD+1, N-1) \left(2-2\cos \frac{\pi}{N+\frac{1}{2}}\right)^{\frac{m}{2}},\nonumber\\
    \geq & 2^m (mD+1)^{mD} e^{-2mD} (N-1)^{-mD} \left(2-2\cos \frac{\pi}{N+\frac{1}{2}}\right)^{\frac{m}{2}},\nonumber\\
    \sim & N^{-m(D+1)}.
    \label{lower_bound_strong}
\end{align}
As a result, we find $z\leq D+1$ as an upper bound of $z$ in the strong-disorder case.

Next, we provide an upper bound of the energy gap.
We use the Rayleigh--Ritz variational principle.
Since the energy spectrum of $\mathbf{M}$ is identical to that of $\mathbf{M}'$
\begin{equation}
    \mathbf{M}'=4(\mathbf{J}_+)^{1-s}\mathbf{R}(\mathbf{J})^{2s}\mathbf{L}(\mathbf{J}_+)^{1-s},
\end{equation}
we obtain from the principle
\begin{align}
   \Lambda_1^2 \leq& (\vec{\psi}, \mathbf{M} \vec{\psi}), \nonumber\\
   \Lambda_1^2 \leq& (\vec{\psi'}, \mathbf{M}' \vec{\psi'}),
   \label{variation}
\end{align}
for $\vec{\psi}$ and $\vec{\psi'}$ satisfying $\|\vec{\psi}\|=1$ and $\|\vec{\psi'}\|=1$.
Here, $(\vec{a}, \vec{b}) \coloneqq \sum_{j=1}^N a_j^* b_j$ is the inner product.

Consider the following two inequalities,
\begin{equation}
    \left\{
    \begin{aligned}
    \mathbf{M} \leq& \tilde{\mathbf{M}} \coloneqq 4(\mathbf{J})^{s}\mathbf{T}(1,1, \Gamma)(\mathbf{J})^{s},\\
    \mathbf{M}' \leq& \tilde{\mathbf{M}}' \coloneqq 4(\mathbf{J}_+)^{1-s}\mathbf{T}(1,1, \Gamma)(\mathbf{J}_+)^{1-s},
    \end{aligned}
    \right.
\end{equation}
These inequalities hold because
\begin{equation}
    \left\{
    \begin{aligned}
    \tilde{\mathbf{M}} - \mathbf{M} =& 4 (\mathbf{J})^s \mathbf{L} [\mathbf{I}-(\mathbf{J}_+)^{2(1-s)}]\mathbf{R} (\mathbf{J})^s\\
    &+ 4 (\mathbf{J})^s [\mathbf{T}(1,1, \Gamma)-\mathbf{T}(0,1, \Gamma)] (\mathbf{J})^s \geq 0,\\
    \tilde{\mathbf{M}}'-\mathbf{M}'=& 4(\mathbf{J}_+)^{1-s}\mathbf{R} [\mathbf{I}-(\mathbf{J})^{2s}] \mathbf{L}(\mathbf{J}_+)^{1-s}\\
    &+ 4 (\mathbf{J})^s [\mathbf{T}(1,1, \Gamma)-\mathbf{T}(1,0, \Gamma)] (\mathbf{J})^s \geq 0.
    \end{aligned}
    \right.
\end{equation}
Here, we have used eq.~(\ref{inequalityT}).
Combining these inequalities with the variational principle gives
\begin{equation}
    \left\{
    \begin{aligned}
    \Lambda_1^2 \leq & (\vec{\psi}, \tilde{\mathbf{M}} \vec{\psi}),\\
    \Lambda_1^2 \leq & (\vec{\psi'}, \tilde{\mathbf{M}}' \vec{\psi'}),
    \end{aligned}
    \right.
\end{equation}
and thus
\begin{equation}
    \left\{
    \begin{aligned}
    \Delta =&\Lambda_1 \leq \sqrt{(\vec{\psi}, \tilde{\mathbf{M}} \vec{\psi})},\\
    \Delta =&\Lambda_1 \leq \sqrt{(\vec{\psi'}, \tilde{\mathbf{M}}' \vec{\psi'})}.
    \end{aligned}
    \right.
    \label{variation_v2}
\end{equation}

We estimate a lower bound of $z$ using the upper bounds of $\Delta$.
We set the vector $\vec{\psi}=(\psi_1, \cdots, \psi_N)$ as
\begin{equation}
    \psi_{j}= (-1)^j C_s J_{j-1}^{-s} \sin \frac{\pi j}{N+1},
\end{equation}
where $J_0 \coloneqq 1$.
The normalization constant $C_s$ is given by
\begin{equation}
    C_s=\left(\sum_{j=1}^N J_{j-1}^{-2s} \sin^2 \frac{\pi j}{N+1}\right)^{-\frac{1}{2}}
\end{equation}
so that $\| \vec{\psi} \|=1$.
Substituting $\vec{\psi}$ and $\Gamma=\Gamma_{\mathrm{c}}=1$ into eq.~(\ref{variation_v2}) gives an upper bound of $\Delta$
\begin{equation}
\Delta \leq \sqrt{(\vec{\psi}, \tilde{\mathbf{M}} \vec{\psi})}=2C_s \sqrt{(N+1) \left( 1 - \cos \frac{\pi}{N+1} \right)}.
\label{upper_bound1}
\end{equation}
Another upper bound is obtained by setting $\vec{\psi'}=(\psi'_1, \cdots, \psi'_N)$ as
\begin{equation}
    \psi'_{j}= (-1)^j C'_s J_{j}^{-(1-s)} \sin \frac{\pi j}{N+1},
\end{equation}
where $J_N \coloneqq 1$ and
\begin{equation}
    C'_s=\left(\sum_{j=1}^N J_{j}^{-2(1-s)} \sin^2 \frac{\pi j}{N+1}\right)^{-\frac{1}{2}}.
\end{equation}
We obtain from eq.~(\ref{variation_v2}) another upper bound of $\Delta$ as
\begin{equation}
    \Delta \leq 2 C'_s \sqrt{(N+1) \left( 1 - \cos \frac{\pi}{N+1} \right)}.
    \label{upper_bound2}
\end{equation}

In both weak-disorder case and strong-disorder case, since $J_j \leq 1$, we obtain
\begin{equation}
    C_s \leq \left( \sum_{j=1}^N \sin^2 \frac{\pi j}{N+1} \right)^{-\frac{1}{2}}=\sqrt{\frac{2}{N+1}}.
\end{equation}
Thus, eq.~(\ref{upper_bound1}) gives for $m \in \mathbb{N}$
\begin{equation}
    [\Delta^m]_{\rm ave} \leq 2^m \left( 2 - 2 \cos \frac{\pi}{N+1} \right)^{\frac{m}{2}}\sim N^{-m}.
    \label{upper_bound_both}
\end{equation}
This indicates $z\ge 1$ as a lower bound of $z$ in both weak-disorder case and strong-disorder case.

In the strong-disorder case, we obtain another lower bound of $z$.
Consider the following inequality,
\begin{align}
    C_s \leq & \left(\sum_{j=\lceil \frac{N+1}{3}\rceil}^{\lfloor \frac{2(N+1)}{3}\rfloor} J_{j-1}^{-2s} \sin^2 \frac{\pi j}{N+1}\right)^{-\frac{1}{2}},\nonumber\\
    \leq & \left(\sin\frac{\pi}{3}\right)^{-1} \left(\sum_{j=\lceil \frac{N+1}{3}\rceil}^{\lfloor \frac{2(N+1)}{3}\rfloor} J_{j-1}^{-2s} \right)^{-\frac{1}{2}},\nonumber\\
    =&\sqrt{3}\left(\sum_{j=\lceil \frac{N+1}{3}\rceil}^{\lfloor \frac{2(N+1)}{3}\rfloor} J_{j-1}^{-2s} \right)^{-\frac{1}{2}},
\end{align}
where $\lceil \cdot \rceil$ and $\lfloor \cdot \rfloor$ denote the ceiling function and floor function, respectively.
Then, eq.~(\ref{upper_bound1}) gives
\begin{align}
    [\Delta^m]_{\mathrm{av}} \leq & 2^m \left[ \left(\sum_{j=\lceil \frac{N+1}{3}\rceil}^{\lfloor \frac{2(N+1)}{3}\rfloor} J_{j-1}^{-2s} \right)^{-\frac{m}{2}} \right]_{\mathrm{av}} \left[3 (N+1) \left( 1 - \cos \frac{\pi}{N+1} \right)\right]^{\frac{m}{2}} \nonumber\\
    \leq & 2^m \left[ \left( \sum_{j=1}^{N_0} J_{j}^{-2s} \right)^{-\frac{m}{2}} \right]_{\mathrm{av}} \left[ 3(N+1) \left( 1 - \cos \frac{\pi}{N+1} \right) \right]^{\frac{m}{2}}, \nonumber\\
    \eqqcolon & 2^m\tilde{C}_{s, N_0}^{(m)} \left[ 3(N+1) \left( 1 - \cos \frac{\pi}{N+1} \right)\right]^{\frac{m}{2}},
    \label{upperbound_v1}
\end{align}
where 
\begin{equation}
N_0 \coloneqq \lfloor \frac{2(N+1)}{3}\rfloor-\lceil \frac{N+1}{3}\rceil +1 \leq \frac{N+5}{3},
\label{Nzero}
\end{equation}
and
\begin{equation}
    \tilde{C}_{s, N_0}^{(m)}=\left[ \left( \sum_{j=1}^{N_0} J_{j}^{-2s} \right)^{-\frac{m}{2}} \right]_{\mathrm{av}}.
\end{equation}
With a method similar to that demonstrated in the above, the other upper bound [eq.~(\ref{upper_bound2})] gives
\begin{equation}
    [\Delta^m]_{\mathrm{av}} \leq 2^m \tilde{C}_{1-s, N_0}^{(m)} \left[ 3(N+1) \left( 1 - \cos \frac{\pi}{N+1} \right)\right]^{\frac{m}{2}}.
    \label{upperbound_v2}
\end{equation}
Then, $\tilde{C}_{s, N_0}^{(m)}$ is bounded by
\begin{align}
    \tilde{C}_{s, N_0}^{(m)}=&N_0! \int_0^1 dJ_1 \int_{J_1}^1 dJ_2 \cdots \int_{J_{N_0-1}}^1 dJ_{N_0} \left(\prod_{i=1}^{N_0} \pi_{\rm s}(J_i) \right)\left(\sum_{j=1}^{N_0} J_j^{-2s}\right)^{-\frac{m}{2}},\nonumber\\
    \leq & N_0! \int_0^1 dJ_1 \int_{J_1}^1 dJ_2 \cdots \int_{J_{N_0-1}}^1 dJ_{N_0} J_1^{mDs} = N_0 B(1+mDs, N_0).
\end{align}
The beta function is bounded by
\begin{align}
    &\ln [N_0 B(1+mDs, N_0)]\nonumber\\
    =&-\sum_{n=1}^{N_0} \ln \left( 1+\frac{mDs}{n}\right),\nonumber\\
    \leq & -\int_1^{N_0+1} \ln \left(1+\frac{mDs}{x}\right) dx,\nonumber\\
    = & (1+mDs) \ln(1+mDs) -(N_0+mDs+1)\ln \left(1+\frac{mDs}{N_0+1}\right) -mDs \ln(N_0+1),\nonumber\\
    \leq & (1+mDs) \ln(1+mDs) -mDs \ln(N_0+1).
\end{align}
Consequently, we obtain
\begin{align}
    \tilde{C}_{s, N_0}^{(m)} \leq& N_0 B(1+mDs, N_0)\nonumber\\
    \leq& (1+mDs)^{1+mDs} (N_0+1)^{-mDs}\nonumber\\
    \leq& (1+mDs)^{1+mDs}\left(\frac{N+8}{3}\right)^{-mDs},
\end{align}
where we have used eq.~(\ref{Nzero}).
We obtain two bounds of $[\Delta^m]_{\mathrm{av}}$ from eqs.~(\ref{upperbound_v1}) and~(\ref{upperbound_v2}) as
\begin{align}
    [\Delta^m]_{\mathrm{av}}\leq& 2^m
    (1+mDs)^{1+mDs} \left(\frac{N+8}{3}\right)^{-mDs}\left[ 3(N+1) \left( 1 - \cos \frac{\pi}{N+1} \right)\right]^{\frac{m}{2}},\nonumber\\
    \sim& N^{-m\left(Ds+\frac{1}{2}\right)},\nonumber\\
    [\Delta^m]_{\mathrm{av}}\leq& 2^m [1+mD(1-s)]^{1+mD(1-s)} \left(\frac{N+8}{3}\right)^{-mD(1-s)}\nonumber\\
    \qquad & \times\left[ 3(N+1) \left( 1 -  \cos \frac{\pi}{N+1} \right)\right]^{\frac{m}{2}},\nonumber\\
    \sim& N^{-m\left(D(1-s)+\frac{1}{2}\right)}.
    \label{upper_bounds}
\end{align}
As a result, in the strong-disorder case, in addition to the bound $z \geq 1$ [see eq.~(\ref{upper_bound_both})], we obtain
\begin{align}
    &z\geq \max\left(Ds+\frac{1}{2}, D(1-s)+\frac{1}{2}\right),\nonumber\\
    &\Leftrightarrow z\geq D\left(\frac{1}{2}+\left|s-\frac{1}{2}\right|\right)+\frac{1}{2}.
\end{align}

We have shown a lower bound and an upper bound of $z$ from the bounds of $[\Delta^m]_{\mathrm{av}}$ in the weak-disorder case [see eqs.~(\ref{lower_bound_weak}) and~(\ref{upper_bound_both})] and in the strong-disorder case [see eqs.~(\ref{lower_bound_strong}),~(\ref{upper_bound_both}) and~(\ref{upper_bounds})].
In the weak-disorder case, since the lower bound and the upper bound of $[\Delta^m]_{\mathrm{av}}$ have the same scaling (i.e., $[\Delta^m]_{\mathrm{av}}\lesssim N^{-m}$ and $[\Delta^m]_{\mathrm{av}}\gtrsim N^{-m}$), we obtain
\begin{equation}
    z=1 \text{ (weak disorder)}.
\end{equation}
This is identical to the case of the transverse field Ising chain without disorder.
Our result is consistent with the numerical result in ref.~\cite{knysh2020quantum}, demonstrating that $z=1$ when $s=1/2$.
On the other hand, in the strong-disorder case, $z$ is not determined by the bounds of $[\Delta^m]_{\mathrm{av}}$.
We obtain
\begin{equation}
    \max \left( D\left(\frac{1}{2}+\left|s-\frac{1}{2}\right|\right)+\frac{1}{2}, 1 \right)\leq z \leq D+1 \text{ (strong disorder)}.
    \label{zbound_strong}
\end{equation}

\subsection{Numerical results}\label{subsec:numerics}
\begin{figure*}[t]
\centering
\begin{tabular}{cc}
	\begin{minipage}{0.5 \linewidth}
	\centering
	\includegraphics[width=\linewidth]{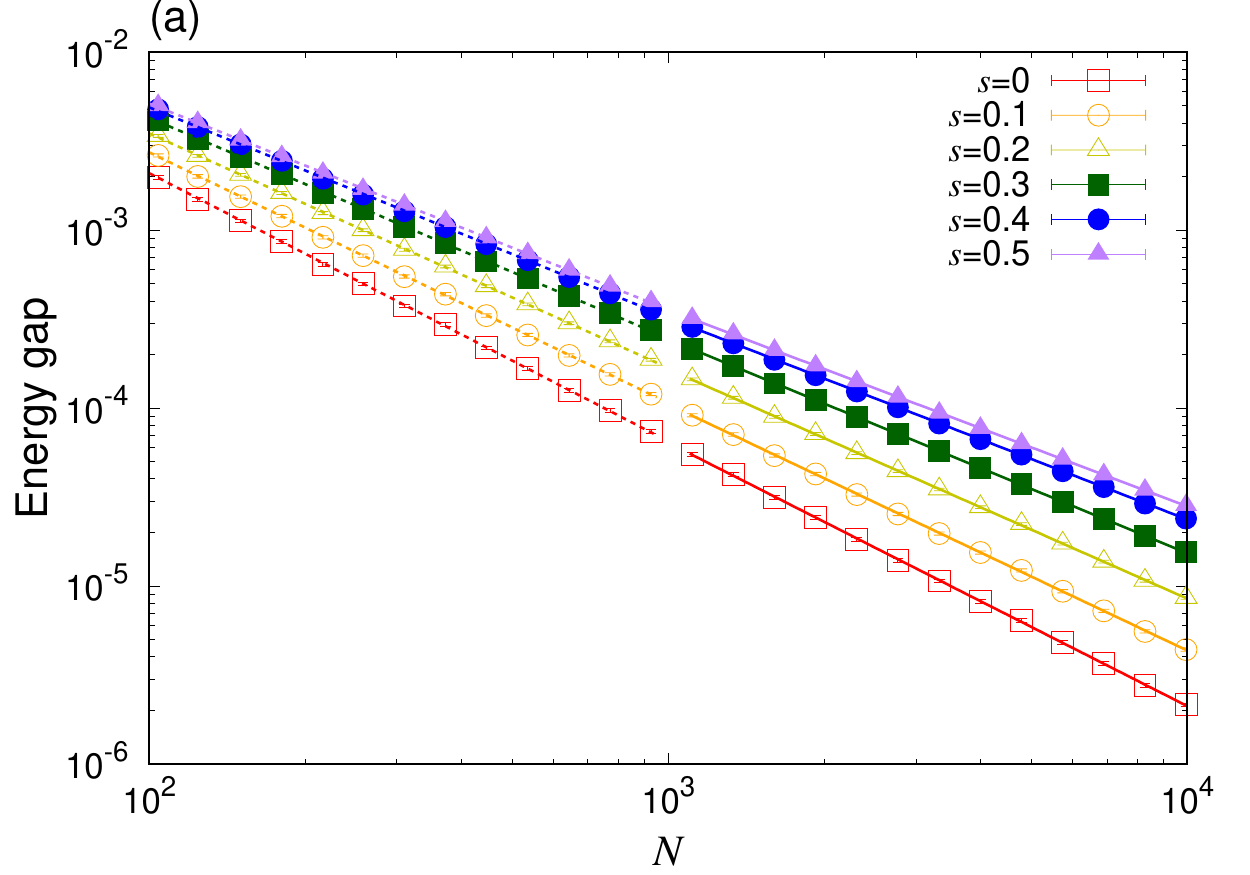}
	\end{minipage}&
	\begin{minipage}{0.5 \linewidth}
	\centering
	\includegraphics[width=\linewidth]{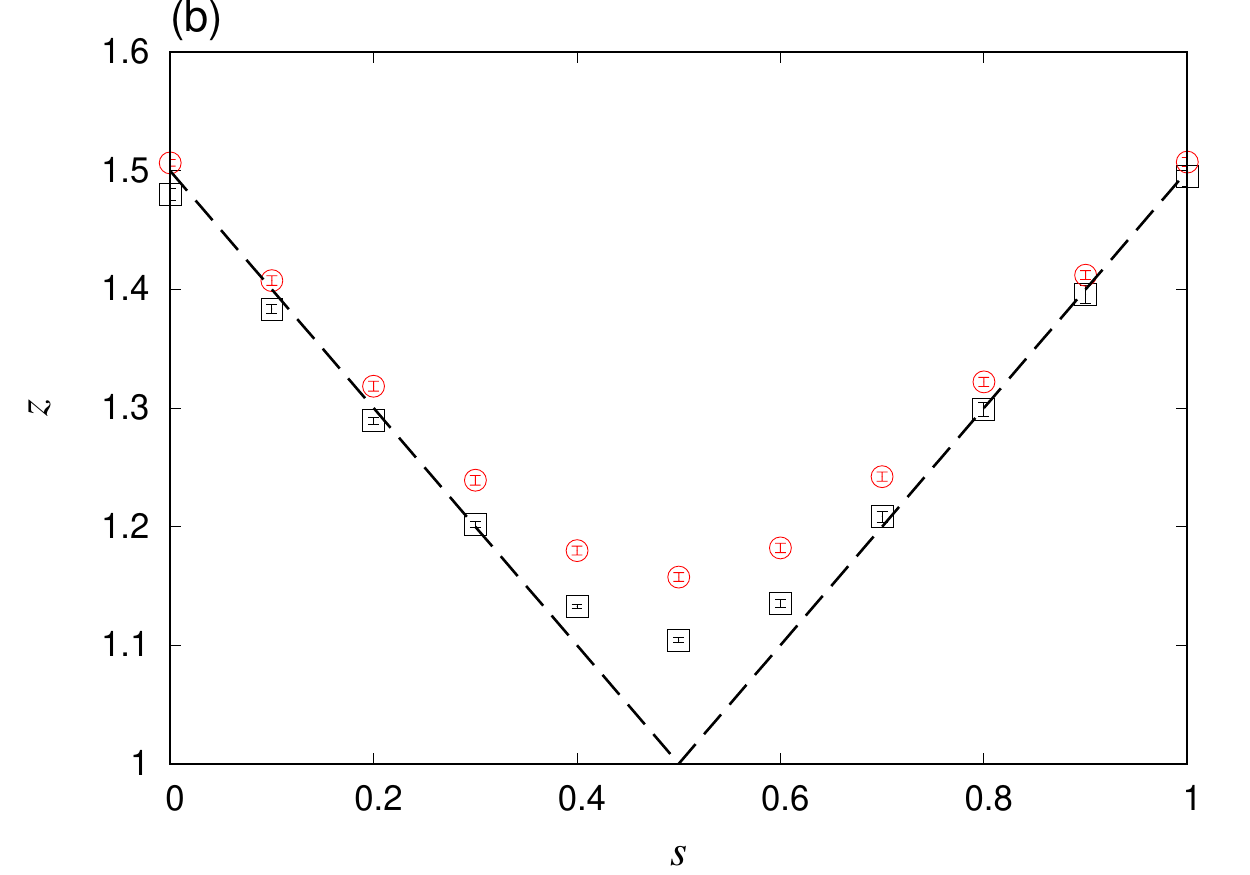}
	\end{minipage}
\end{tabular}
\caption{
(Color online)
(a) System-size dependences of the energy gap $[\Delta]_{\mathrm{av}}$ at $D=1$ for various values of $s$ [$s=0$ (open squares), $s=0.1$ (open circles), $s=0.2$ (open triangles), $s=0.3$ (filled squares), $s=0.4$ (filled circles), $s=0.5$ (filled triangles)].
Data for $s=0.6, 0.7, 0.8, 0.9,$ and $1$ are omitted since the data for $s$ are overlapped with those for $1-s$.
The bold lines and the dotted lines are obtained by fitting the data for $N\ge 10^3$ and the data for $10^2 \leq N \leq 10^3$, respectively.
Each data point is obtained by averaging $1000$ realizations of disorder.
(b) $s$-Dependence of the dynamical critical exponent $z$.
$z$ is estimated by the regression analysis in Fig.~(a) and the error bars denote the standard deviation for the slopes of the regression lines.
Squares and Circles denote $z$ obtained by using the data for $N\ge 10^3$ and the data for $10^2 \leq N \leq 10^3$, respectively.
Dotted lines give the lower bound of $z$ [see eq.~(\ref{zbound_strong})].
}
\label{fig_random}
\end{figure*}

This subsection gives the numerical results of $z$ in the strong-disorder case.
An exact diagonalization of the matrix $\mathbf{M}$ is performed to obtain $[\Delta]_{\mathrm{av}}$.

Figure~\ref{fig_random}(a) shows the $N$-dependences of $[\Delta]_{\mathrm{av}}$ at $D=1$ for various values of $s$.
Each $s$ has a power-law scaling.
The power corresponds to $z$, which is numerically obtained by fitting the data.
Figure~\ref{fig_random}(b) shows the $s$-dependence of $z$.
In figure, $z$ estimated by fitting the data for $N\ge 10^3$ and $10^2 \leq N \leq 10^3$ are depicted by squares and circles, respectively.
$z$ at $s$ and $1-s$ are the same and $z$ at $s=0.5$ is the smallest.
When $s \ge 0.7$ and $s \le 0.3$, $z$ is well described by the lower bound of $z$ [see eq.~(\ref{zbound_strong})].
Some $z$ are slightly smaller than the lower bound of $z$.
We expect that the reason for this discrepancy is due to finite-size effects.
Large finite-size effects are observed around $s \simeq 0.5$, and thus the study of larger system sizes is necessary to determine the values of $z$ in this regime. 

\section{Conclusion and outlook} \label{sec:conclusion}
We investigated the effect of a correlated disorder in a transverse-field Ising chain on the QPT properties.
We obtained the lower and upper bounds of $[\Delta^m]_{\mathrm{av}} (m \in \mathbb{N})$, and we analytically showed that $z$ is finite; $z=1$ in the weak-disorder case and $\max (D(1/2+|s-1/2|)+1/2, 1)\leq z \leq D+1$ in the strong-disorder case, where $s$ and $D$ are parameters for the tuning the transverse fields and the disorder strength, respectively.

In the strong-disorder case we numerically showed that $z$ at $D=1$ depends on $s$ and $z$ is well described by the lower bound of $z$: $z=1+|s-1/2|$ for $s \le 0.3$ and $s \ge 0.7$.
It is challenging in mathematics to analytically obtain the dependences of $z$ on $s$ and $D$ from the matrix $\mathbf{M}$ in eq.~(\ref{matrixM}).

This study shows that $z$ is lowered from infinity to a finite value by tuning the transverse field strength.
The suppression of $z$ is useful to enhance the performance of adiabatic quantum computations~\cite{kibble1976topology,zurek1985cosmological,zurek2005dynamics, dziarmaga2010dynamics, polkovnikov2011colloquium, delcampo2014universality}.
Therefore, it is important to lower $z$ in more general quantum many-body systems.
Future work includes studying the effect of a tuned inhomogeneous transverse field on the model in $d$-dimension ($d \ge 2$)~\cite{rieger1996griffiths,guo1996quantum,pich1998critical,nishimura2020griffiths}, where the spin-glass phase can appear due to the presence of frustration.

\section*{Acknowledgments}
T.~S. thanks Kensuke Tamura, Sei Suzuki, and Jos{\'e} A Hoyos for the fruitful discussions and comments.
The authors are very grateful to Hosho Katsura for helping us improve the presentation, especially that for the lower bound of $\Delta$ (the upper bound of $z$).
T.~S. was partially supported by JSPS KAKENHI (Grant Number 18K13466).
S.~T. was partially supported by JSPS KAKENHI (Grant Number 19H01553).
This paper is partially based on the results obtained from a project commissioned by the New Energy and Industrial Technology Development Organization (NEDO).
T.~S. and S.~T. thank the Supercomputer Center, the Institute for Solid State Physics, The University of Tokyo, and the Yukawa Institute for Theoretical Physics for the use of the facilities.

\appendix

\section{A fermionic view of the eigenvalue problem for the energy gap and the correlation function}~\label{appendixA}
Here, details to calculate the energy gap and the correlation function are shown using a transformation of a spin system to a free fermion system.
In the Jordan--Wigner transformation~\cite{lieb1961two},
the fermion operator is described by
\begin{equation}
    \left\{
    \begin{aligned}
    c_j&=-\frac{\mathrm{i}}{2}\prod_{k=1}^{j-1} (-\sigma_k^x) (\sigma_j^y-{\mathrm{i}}\sigma_j^z),\\
    c_j^\dagger&=\frac{\mathrm{i}}{2}\prod_{k=1}^{j-1} (-\sigma_k^x) (\sigma_j^y+{\mathrm{i}}\sigma_j^z),
    \end{aligned}
    \right.
\end{equation}
where $c_j$ and $c_j^\dagger$ are the annihilation and the creation operator of the fermion at site $j$, respectively.
The Hamiltonian in eq.~(\ref{eq:model_intro}) in the main text is described in a quadratic form of fermion operators as
\begin{equation}
    H=-\sum_{i=1}^{N-1} J_i (c_{i}^\dagger - c_{i})(c_{i+1}^\dagger + c_{i+1})-\sum_{i=1}^N \Gamma_i (2 c_i^\dagger c_i -1).
\end{equation}
When matrices $\mathbf{A}$ and $\mathbf{B}$ are introduced, the Hamiltonian is written as
\begin{equation}
    H=\sum_{i=1}^N \sum_{j=1}^N \left[ c_i^\dagger [\mathbf{A}]_{ij} c_i +\frac{[\mathbf{B}]_{ij}}{2}(c_i^\dagger c_j^\dagger +c_j c_i) \right]+\text{const.},
    \label{hamil_f}
\end{equation}
where
\begin{equation}
\left\{
\begin{aligned}
& [\mathbf{A}]_{i,i}=-2\Gamma_i, \nonumber\\
&[\mathbf{A}]_{i,i+1}=-J_i, [\mathbf{A}]_{i+1,i}=-J_i, \text{ for } 1 \leq i \leq N-1, \nonumber\\
&\text{otherwise } 0,
\end{aligned}
\right.
\end{equation}
and
\begin{equation}
\left\{
\begin{aligned}
&[\mathbf{B}]_{i, i+1}=-J_i, [\mathbf{B}]_{i+1, i}=J_i \text{ for } 1 \leq i \leq N-1, \nonumber\\
&\text{otherwise } 0.
\end{aligned}
\right.
\end{equation}
The matrix $\mathbf{A}$ is an a symmetric matrix and $\mathbf{B}$ is an alternative matrix.

Since the Hamiltonian is written in the quadratic form, it can be transformed into the following diagonal form
\begin{equation}
    H=\sum_{k=1}^N \Lambda_k \eta_k^\dagger \eta_k+\text{const.},
    \label{diagonal}
\end{equation}
using the Bogoliubov transformation,
\begin{align}
    \eta_k=&\sum_{i=1}^N (g_{k,i}c_i+h_{k,i}c_i^\dagger), \nonumber\\
    \eta_k^\dagger =& \sum_{i=1}^N (g_{k,i}c_i^\dagger+h_{k,i}c_i),
    \label{Bogoliubov}
\end{align}
where $g_{k,i}, h_{k,i}\in \mathbb{R}$.
Here, $\eta_k$ and $\eta_k^\dagger$ satisfy the anticommutation relation of fermions, and $\Lambda_k (\geq 0)$ is regarded as a quasienergy of a quasiparticle.
The quasienergy $\Lambda_k$ is arranged in ascending order (i.e., $0 \leq \Lambda_1 \leq \Lambda_2 \leq \cdots$).
The coefficients $g_{k,i}$ and $h_{k,i}$ satisfy
\begin{align}
    \Lambda_k g_{k,i} &= \sum_{j=1}^N ([\mathbf{A}]_{ij} g_{k,j} +[\mathbf{B}]_{ij} h_{k,j}),\nonumber\\
    \Lambda_k h_{k,i} &= -\sum_{j=1}^N ([\mathbf{B}]_{ij} g_{k,j} +[\mathbf{A}]_{ij} h_{k,j}).
\end{align}
In terms of the new variables,
\begin{equation}
    \phi_{k,i}=g_{k,i}+h_{k,i}, \quad \psi_{k,i}=g_{k,i}-h_{k,i},
\end{equation}
the coupled equations are
\begin{align}
\sum_{j=1}^N [\mathbf{A}+\mathbf{B}]_{ij} \phi_{k,j}&=\Lambda_k \psi_{k,i},\nonumber\\
\sum_{j=1}^N [\mathbf{A}-\mathbf{B}]_{ij} \psi_{k,j}&=\Lambda_k \phi_{k,i}.
\label{coupled}
\end{align}
Then, the problem is reduced to the following eigenvalue problem~\cite{lieb1961two,schultz1964two}
\begin{equation}
    \sum_{j=1}^N [\mathbf{M}]_{ij}\phi_{k,j} \coloneqq \sum_{j=1}^N [(\mathbf{A}-\mathbf{B})(\mathbf{A}+\mathbf{B})]_{ij}\phi_{k,j}=\Lambda_k^2 \phi_{k,i},
    \label{eigenvalue}
\end{equation}
where $\mathbf{M}=(\mathbf{A}-\mathbf{B})(\mathbf{A}+\mathbf{B})$.
For $\Lambda_k \neq 0$, eq.~(\ref{eigenvalue}) is solved for $\{\phi_{k,i}\}_{i=1}^N$, and $\{ \psi_{k,i} \}_{i=1}^N$ are obtained from
\begin{equation}
    \psi_{k,i}=\frac{1}{\Lambda_k} \sum_{j=1}^N [\mathbf{A}+\mathbf{B}]_{ij} \phi_{k,j}.
\end{equation}
For $\Lambda_k = 0$, $\{\phi_{k,i}\}_{i=1}^N$ and $\{ \psi_{k,i} \}_{i=1}^N$ are determined by Eq.~(\ref{coupled}), and their relative sign is arbitrary.

The energy gap in the language of fermions is obtained by
\begin{equation}
    \Delta = \Lambda_1.
    \label{gap}
\end{equation}
We have used it in the main text to analytically and numerically evaluate dynamical critical exponents.

The correlation function in the ground state $C_{i,j}$ for $j>i$ is given by~\cite{young1996numerical, lieb1961two}
\begin{equation}
    C_{i,j}\coloneqq \langle \sigma_i^z \sigma_j^z \rangle_{\rm g}=\det \mathbf{G}_{i,j},
    \end{equation}
where $\langle \cdot \rangle_{\rm g}$ denotes the expectation value in the ground state.
$\mathbf{G}_{i,j}$ is a matrix with $j-i$ dimension and its matrix element is given by
\begin{align}
    [\mathbf{G}_{i,j}]_{kl}=&\langle (c_{i+k-1}^\dagger -c_{i+k-1})(c_{i+l}^\dagger +c_{i+l}) \rangle_{\rm g}, \nonumber\\
    =&-\sum_{n=1}^N \psi_{n,i+k-1} \phi_{n,i+l}
\end{align}
The labels $k$ and $l$ run from $1$ to $j-i$.

\section{Average and typical correlation lengths}~\label{appendixB}
\begin{figure}[t]
\centering
	\includegraphics[width=\linewidth]{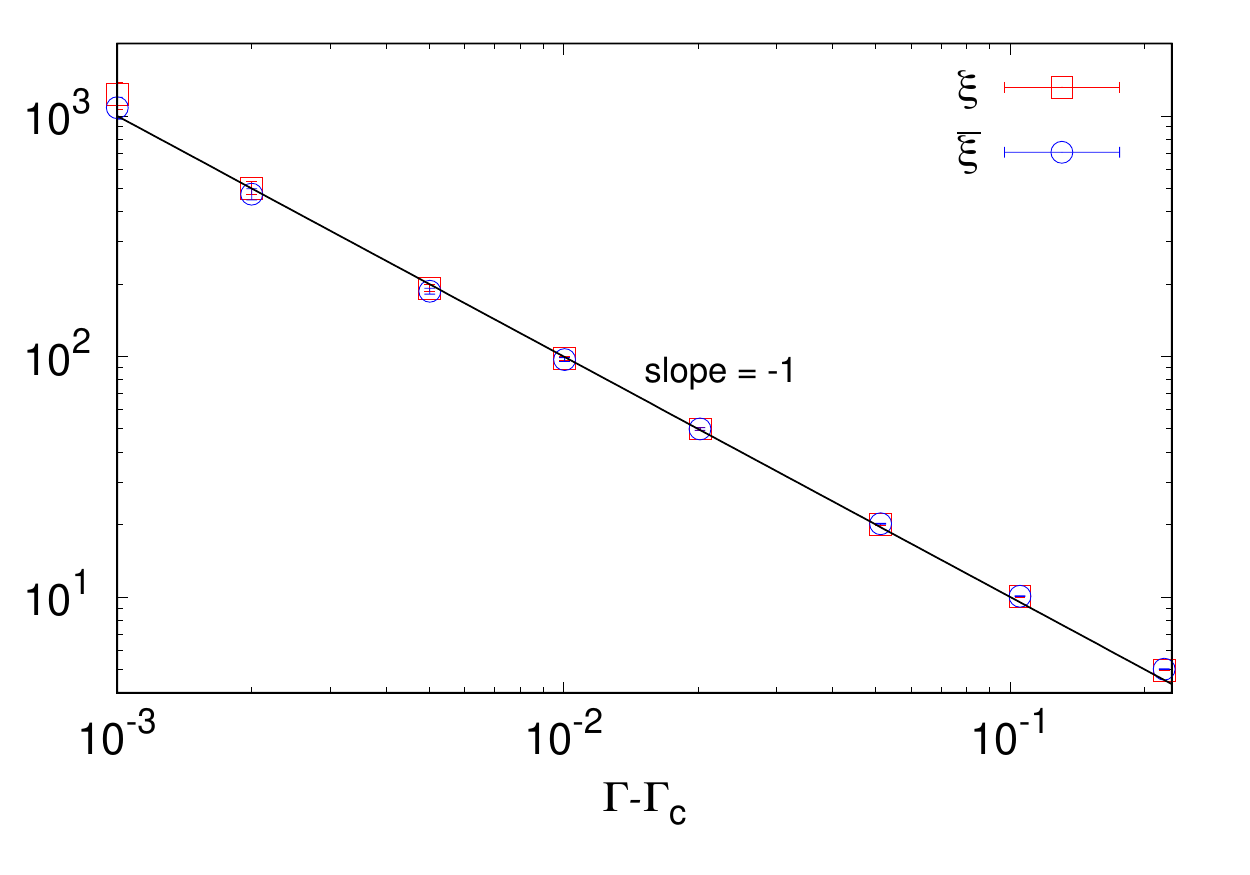}
\caption{
(Color online)
Average and typical correlation lengths $\xi$ and $\bar{\xi}$ as a function of $\Gamma-\Gamma_{\rm c}$ in the strong-disorder case.
The parameters $N=8000$, $D=1$, and $s=0.5$ are used.
The error bars denote the standard deviation for fitting the correlation functions, which are obtained by averaging $20$ realizations of disorder.
Both correlation lengths are well described by the bold line with slope $-1$ in the log-log plot (i.e., $\nu=\bar{\nu}=1$).
}
\label{fig_correlation}
\end{figure}
This appendix numerically compares the average and typical correlation lengths in the transverse-field Ising chain with a correlated disorder.
The correlation lengths are obtained by the average correlation function $C_{\rm ave}(x)$ and the typical correlation function $C_{\rm typ}(x)$,
\begin{equation}
    \left\{
    \begin{aligned}
    &C_{\rm ave}(x) \coloneqq [C_{N/2,N/2+x}]_{\rm ave},\\
    &\ln C_{\rm typ}(x) \coloneqq [\ln C_{N/2,N/2+x}]_{\rm ave},
    \end{aligned}
    \right.
\end{equation}
where $C_{i,j}$ is herein calculated by using an exact diagonalization method (see~\ref{appendixA}).
At the critical point, the correlation functions vary as a power of $x$,
\begin{equation}
    \left\{
    \begin{aligned}
    &C_{\rm ave}(x) \sim x^{-\eta},\\
    &C_{\rm typ}(x) \sim x^{-\bar{\eta}},
    \end{aligned}
    \right.
\end{equation}
where $\eta$ and $\bar{\eta}$ are critical exponents.
Away from the critical point in the paramagnetic phase, the correlation functions decay exponentially as
\begin{equation}
    \left\{
    \begin{aligned}
    &C_{\rm ave}(x) \sim \exp \left(-\frac{x}{\xi}\right),\\
    &C_{\rm typ}(x) \sim \exp \left(-\frac{x}{\bar{\xi}}\right),
    \end{aligned}
    \right.
\end{equation}
where $\xi$ and $\bar{\xi}$ are the average and typical correlation lengths.
The diverging length scale is characterized by the critical exponents $\nu$ and $\bar{\nu}$ as
\begin{equation}
    \left\{
    \begin{aligned}
    &\xi \sim (\Gamma-\Gamma_{\rm c})^{-\nu},\\
    &\bar{\xi} \sim (\Gamma-\Gamma_{\rm c})^{-\bar{\nu}}.
    \end{aligned}
    \right.
\end{equation}
In the uncorrelated disorder cases, the correlation function $C_{i, j}$ has large sample-to-sample fluctuations, and as a result, $\nu$ and $\bar{\nu}$ take different values ($\nu=2$ and $\bar{\nu}=1$)~\cite{young1996numerical, fisher1995critical, fisher1992random}.

Figure~\ref{fig_correlation} shows the dependences of $\xi$ and $\bar{\xi}$ on $\Gamma$ in the strong-disorder case.
The parameters are set as $(N, D, s)=(8000, 1, 0.5)$.
The correlation lengths $\xi$ and $\bar{\xi}$ are calculated by fitting the correlation functions $C_{\rm ave}(x)\sim \exp (-x/\xi) x^{-\eta}$ and $C_{\rm typ}(x)\sim \exp (-x/\bar{\xi}) x^{-\bar{\eta}}$.
Unlike the uncorrelated disorder cases, $\xi$ and $\bar{\xi}$ are overlapped with each other and the critical exponents are given by
\begin{equation}
    \nu=\bar{\nu}=1.
\end{equation}
$\nu=1$ is consistent with ref.~\cite{hoyos2011protecting}.

\bibliography{QA}

\end{document}